\documentclass[12pt,epsfig,citesort,svgnames,dvipsnames]{article}
\usepackage{multicol}
\usepackage[all]{xy}
\usepackage{graphicx}
\usepackage{todonotes}
\usepackage{amsmath}
\usepackage{amssymb}
\usepackage{amsfonts}
\usepackage{amsthm}
\newtheorem{ten}{Tenet}

\newtheorem{tenn}{Revised Tenet}

\usepackage{amscd}
\usepackage{latexsym}
\usepackage[official]{eurosym}
\usepackage[english]{babel}
\usepackage[T1]{fontenc}
\usepackage[utf8]{inputenc}
\usepackage{authblk}
\usepackage{natbib}
\usepackage{xcolor}
\usepackage{hyperref}
\usepackage[nottoc]{tocbibind}
\hypersetup{
colorlinks=true,
citecolor=WildStrawberry,
anchorcolor=Black,
filecolor=cyan,
menucolor=magenta,
runcolor=cyan,
linkcolor=RoyalBlue,
linktoc=page, 
urlcolor=TealBlue}

\title{\bf {Shape space as a conceptual space}}
\author{\normalsize Antonio Vassallo}
\affil{\normalsize \emph{Warsaw University of Technology, Faculty of Administration and Social Sciences, Plac Politechniki 1, 00-661 Warsaw, Poland}\\antonio.vassallo1977@gmail.com}

\date{}

\begin{document}
\maketitle

\begin{center}
Forthcoming in \emph{Synthese}.
\end{center}

\begin{abstract}
The notion of shape space was introduced in the second half of the 20th Century as a useful analytical tool for tackling problems related to the intrinsic spatial configuration of material systems. In recent years, the geometrical properties of shape spaces have been investigated and exploited to construct a totally relational description of physics (classical, relativistic, and quantum). The main aim of this relational framework---originally championed by Julian Barbour and Bruno Bertotti---is to cast the dynamical description of material systems in dimensionless and scale-invariant terms only. As such, the Barbour-Bertotti approach to dynamics represents the technical implementation of the famous Leibnizian arguments against the reality of space and time as genuine substances. The question then arises about the status of shape space itself in this picture: Is it an actual physical space in which the fundamental relational dynamics unfolds, or is it just a useful mathematical construction? The present paper argues for the latter answer and, in doing so, explores the possibility that shape space is a peculiar case of a conceptual space.
 \\
    \\
\textbf{Keywords}: Shape space; conceptual space; Leibnizian/Machian relationalism; pure shape dynamics.
\end{abstract}

\tableofcontents

\section{Introduction: Space and time as keys to physical knowledge}\label{sec:int}

Space and time are so deeply entrenched in the way we conceive and access the physical world that they have been aptly called ``touchstones of reality'' since ``[e]verything real is somewhere, sometime'' \citep[][\S~1.2]{1}. This intuition rests on two commitments: An ontic one---physical systems exist in space and time---and an epistemic one---physical systems are known through their spatiotemporal representation. It is then no surprise that the doctrine that fully embraces these two commitments---\emph{substantivalism}---has dominated the philosophical landscape for more than three centuries, i.e., since Newton laid down its fundamental principles in the \emph{Scholium} to his \emph{Principia}. 

Indeed, classical mechanics, as formulated by Newton, describes physical systems in terms of absolute space and time. These immutable structures serve as the stage in which material bodies move and interact, providing fixed references for positions, velocities, and accelerations. In modern times, Einstein's theory of general relativity advanced a significant shift by treating spacetime not as a static backdrop but as a dynamic entity influenced by material fields. Nevertheless, even in general relativity, spacetime persists as a kind of global structure that underpins physical processes.

Substantivalism, in any of its versions, is a very appealing position in that it provides powerful explanations not only about the way physical bodies move, but also about the way we get to experience the physical world. In fact, by accepting that space and time are self-subsisting entities, we immediately realize why they constitute the hallmarks of what is physical (because the physical world is inherently spatial and temporal) and why they render the physical world intelligible to our minds (because cognitive abilities are the product of evolutionary processes shaped by an inherently spatial and temporal external environment). 

Given that substantivalism purportedly delivers a doubly firm grasp on reality (ontic \emph{and} epistemic), it is obvious that challenging this philosophical doctrine is an arduous task. This endeavor was famously taken up by Leibniz, who articulated the shift arguments against the reality of space and time in his correspondence with Samuel Clarke \citep{717}. Although these arguments have a diverse and nuanced structure (see, for example, \citealp{17}, \S~3), they all go in the direction of highlighting how a commitment to substantival space and time leads to observationally identical yet ontologically distinct states of affairs that lack a sound justification for their existence. 

For example, in the ``static'' version of these arguments, Leibniz reasoned that if space were an independent, self-subsisting entity, then it would be meaningful to imagine shifting the entire universe by a certain distance within this absolute space. However, because the relative positions and distances among all material objects would remain unchanged, such a shift would result in no observable difference whatsoever. Leibniz concluded that positing an absolute space that allows for physically indistinguishable but supposedly distinct scenarios introduces unnecessary metaphysical baggage.

The bottom line of Leibniz's shift arguments is that substantivalism entails a commitment to much more structure than needed to account for empirical observations and, hence, a more parsimonious ontology should be sought, which prevents these empirically superfluous states of affairs from arising. The alternative ontology Leibniz proposed in view of these arguments consists of spatial relations among ``coexisting'' material bodies and their mutual change, or ``order of succession.'' This is what is nowadays known as \emph{Leibnizian relationalism}. This doctrine is also called \emph{Machian relationalism}, acknowledging the later insights provided by the German physicist Ernst Mach \citep[][\S~6.2]{464}. 

Relationalism as a philosophical doctrine has for a long time suffered from the opposite problem faced by substantivalism: It was charged with postulating too meager an ontology to ground (classical) dynamics. An influential argument in this sense comes from Newton's famous bucket experiment. In this thought experiment, Newton considered a bucket filled with water, suspended by a rope. When the bucket is spun, the surface of the water becomes concave. Importantly, the concavity of the water's surface persists even when the bucket and water rotate uniformly together, eliminating any relative motion between them. Newton took this as evidence that the concavity reflects the water's rotation relative to absolute space, not relative to other bodies. This argument suggests that certain physical effects, such as inertial forces, require an absolute frame of reference, challenging the sufficiency of relational descriptions that rely solely on the relative positions and motions of bodies.

In a similar vein, the French mathematician and physicist Henri Poincar\'e famously pointed out how the angular momentum of a globally rotating system of particles cannot be fixed in Newtonian dynamics given only the instantaneous inter-particle separations and their rate of change \citep{poi}. This, in particular, means that any theoretical implementation of Leibnizian/Machian relationalism in a classical setting is unable to recover the full set of empirical predictions that can be formulated in Newtonian mechanics---case in point, physical situations in which the whole material content of the universe rotates (i.e., where the total angular momentum $\mathbf{J}$ is non-zero).

In light of what we have already said, the problems faced by Leibnizian/Machian relationalism also have an epistemic dimension. How can we make good epistemic sense of an ontology that denies the existence of full-fledged spatial and temporal structures if, in the end, material bodies become intelligible to us by virtue of them being placed in space at a certain time? In short, Leibnizian/Machian relationalism seems to deny that space and time are touchstones of reality without supplying a consistent story about why phenomenal experience appears to have essential spatial and temporal connotations. It took several centuries and the work of many physicists and philosophers to come up with a modern Leibnizian/Machian paradigm that stands a chance to avoid these issues (see \citealp{77}, for a historical overview of the conceptual development of relationalism). The birth of modern relational dynamics can be traced back to the seminal work of Julian Barbour and Bruno Bertotti \citep{83}, which eventually led to the theory known as \emph{shape dynamics} \citep[][is a comprehensive textbook on this subject]{514}. The latest evolution of this Leibnizian/Machian framework is \emph{pure shape dynamics} \citep{746}. The paper will focus on this latter framework and consider how it accounts for the intelligibility of physical dynamics in the absence of underlying spatial and temporal structures. The key role in this sense will be played by \emph{shape space}, the relational equivalent of standard configuration space.  

The plan of the paper is as follows: \S~\ref{sec:one} will briefly present the technical framework of pure shape dynamics, showing how shape space arises in this context; the section will argue that shape space is the relational surrogate of the spatiotemporal ``arena'' postulated by substantivalism. \S~\ref{sec:two} will address the issue of how a theory like pure shape dynamics, which dispenses with space and time as fundamental entities, can still render the dynamics of physical objects intelligible to the human mind; the question will be considered as to whether modern Leibnizian/Machian relationalism requires a double commitment (ontic and epistemic) to shape space, and it will be argued that only the epistemic commitment is necessary to make sense of the dynamics: This is where the notion of conceptual space will be invoked and its relation to shape space will be discussed. Finally, \S~\ref{sec:disc} will deliver a few concluding remarks, addressing in particular how the defended thesis can be carried over to relativistic and quantum physics.

\section{No time, no space, just shapes}\label{sec:one}

The genesis of modern relational dynamics can be traced back to the work of Julian Barbour and Bruno Bertotti \citep{720, 84, 83}. These authors wanted to codify in a formal framework the relationalist's ambition to eliminate redundant structures from physical theories---i.e., those structures that typically emerge due to the presence of various symmetries in the system under investigation. This drive to streamline physical theories aligns with an empiricist perspective, according to which all structures whose variation results in no empirically observable difference should be excluded from physics. Not surprisingly, Newton's concepts of absolute space and time, with their questionable causal efficacy and unobservable nature, are prime candidates for elimination.

Barbour and Bertotti's original input led to the formulation of what is today known as the \emph{Mach-Poincar\'e principle} (see \citealp{730}, \S~4.2, for a full discussion of the relationalist tenets):

\begin{ten}[\textbf{Mach-Poincar\'e Principle -- Classic Version}]
\label{MachP}
Physical, i.e., relational initial configurations and their (intrinsic) first derivatives alone should uniquely determine the dynamical evolution of a closed system.
\end{ten}

Note how this principle is underpinned by two strong anti-substantivalist theses:

\begin{ten}[\textbf{Spatial Relationalism -- Classic Version}]
\label{MachP2}
Lengths, be they distances or sizes, must be defined relative to physical systems, not spatial points.
\end{ten}

\begin{ten}[\textbf{Temporal Relationalism}]
Temporal structures, such as chronological ordering, duration, and temporal flow, must be defined only in terms of changes in the relational configurations of physical systems.
\end{ten}

A key emphasis for the modern relationalist is that all measurements are fundamentally comparisons between physical systems. Formally, this means that only \emph{ratios} of physical quantities carry objective information. In spatial terms, this comparativist view implies that we should dispense with the notion of absolute size, focusing instead on the \emph{shape} of systems. Consider, for example, three particles in Newtonian space positioned at the vertices of a triangle. Let $x_a$ represent the particle positions and $r_{ab}\equiv|x_a-x_b|$ the inter-particle separations. The relational view dictates that one of the $r_{ab}$ should serve as a unit of length, with the remaining distances expressed as ratios relative to this unit. This process reveals that only the shape of the triangle matters, characterized by two independent angles (i.e., its conformal structure). This leads us to a modern reformulation of spatial relationalism:

\setcounter{tenn}{1}

\begin{tenn}[\textbf{Spatial Relationalism -- Modern Version}]\label{MachP3}
The only physically objective spatial information of a system is encoded in its shape, understood as its dimensionless and scale-invariant relational configuration.
\end{tenn}

\begin{figure}[h!]
\begin{center}
\includegraphics[scale=0.6]{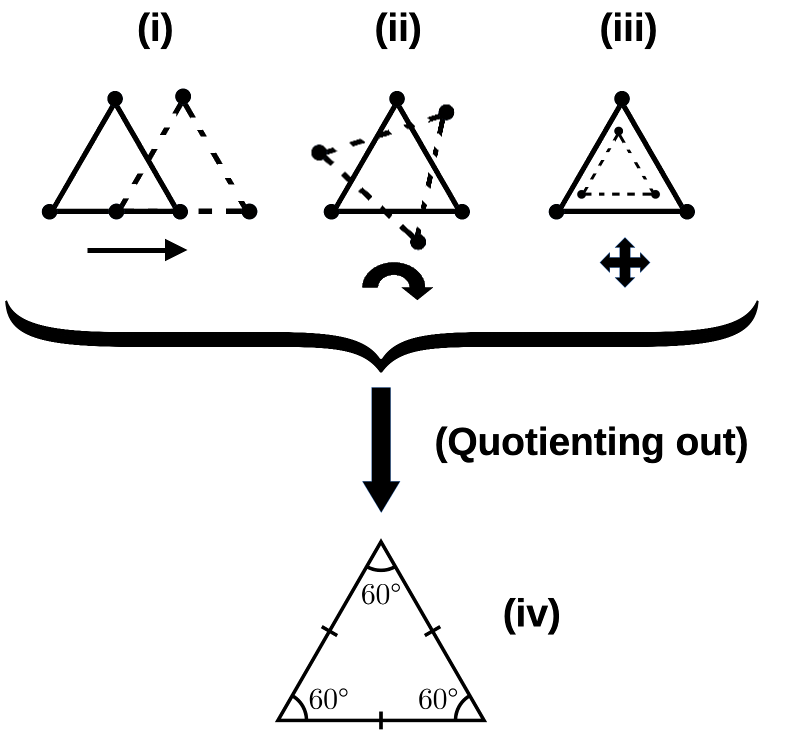}
\caption{In the simple case of three particles arranged in an equilateral triangle configuration placed in a two-dimensional Euclidean space, the quotienting out procedure removes all degrees of freedom associated with rigid translations (i), planar rotations (ii), and scalings (iii). What is left at the end of the procedure is the conformal structure (internal angles) that uniquely characterizes the shape of the configuration (iv). From a Leibnizian/Machian perspective, this procedure implements the idea that there really is no space in which the three particles are placed---the only relevant physical information is encoded in the way they are mutually related.}
\label{fig0}
\end{center}
\end{figure}

Mathematically, the process of eliminating redundant structures associated with symmetry is known as \emph{quotienting out}. If $\mathcal{Q}$ represents the relevant configuration space of a system and $\mathcal{G}$ the symmetry group, the physically meaningful configuration space is $\mathcal{Q}_{ss}:=\mathcal{Q}/\mathcal{G}$, with dimension $\mathrm{dim} (\mathcal{Q}_{ss})=\mathrm{dim}(\mathcal{Q})-\mathrm{dim}(\mathcal{G})$. For $N$ classical particles, $\mathcal{Q}$ is the standard configuration space, and $\mathcal{G}=\mathsf{Sim(3)}$, the similarity group comprising Euclidean translations ($\mathsf{T}$), rotations ($\mathsf{R}$), and dilations ($\mathsf{S}$)---these latter being scale transformations, which alter the size of a configuration (figure \ref{fig0}). The resulting quotient space, $\mathcal{Q}_{ss}=\mathcal{Q}/\mathsf{Sim(3)}$, is the \emph{shape space} of the system, wherein each point corresponds to an $N$-gon's shape (i.e., an equivalence class of $N$-gons under $\mathsf{T,R,\text{and},S}$ transformations; see \citealp{449}, for a comprehensive textbook on shape spaces).

Barbour and Bertotti effectively implemented tenet \ref{MachP} in the context of classical particle dynamics via \emph{best-matching}, an intrinsic derivative that provides a measure of \emph{equilocality}---the ability to determine when two points are identically positioned at different times based solely on relational data. However, recalling Poincar\'e's work mentioned earlier, observable initial configurations and their first derivatives alone do not suffice to determine a system's evolution due to the total angular momentum $\mathbf{J}$. Best-matching results in the vanishing of $\mathbf{J}$ for a \emph{closed} system, like the whole universe, though subsystems may still possess non-zero angular momentum. Thus, fully relational dynamics apply uniquely to closed systems. This is, in a nutshell, the crucial holistic nature of relational physics, as it was envisioned by Mach with his ``non-local'' explanation of the origin of inertia in terms of what is nowadays known as \emph{frame-dragging} effect \citep[see, e.g., the discussion in][]{557}. The pursuit of a gravity theory adhering to the relational principles of Barbour and Bertotti eventually culminated in the already mentioned shape dynamics (SD), further advancing the relational program in modern physics \citep{528}.


The latest refinement in this ongoing relational program is pure shape dynamics (PSD; see \citealp{746}, for an in-depth technical introduction). In essence, ``Pure'' in PSD signifies a description of any dynamical theory purely through the intrinsic geometric properties of the \emph{unparametrized} curve $\gamma_0$ traced by the physical system in shape space $\mathcal{Q}_{ss}$. This approach ensures that no reference structures or clock processes external to the system are needed to describe $\gamma_0$ in $\mathcal{Q}_{ss}$. In contrast, the original SD requires a monotonically increasing parameter---such as the ratio of dilatational momenta or York time/spatial volume---to make sense of physical evolution. This parameter, although weaker than Newtonian time, remains an external element to the system \citep[cf.][\S~2.2, for a detailed comparison between SD and PSD]{732}. Thus, PSD is an innovative project aimed at rewriting the entire spectrum of dynamical theories, both historical and future, in fully relational, scale-invariant terms. For any ``standard'' physical system, such as a Newtonian $N$-particle system, a general relativistic cosmological model, or a de Broglie-Bohm particle system \citep{fark}, PSD applies the same method: Eliminate the relevant symmetries from the system and then completely geometrize its dynamical development.

This insistence on the intrinsic geometric properties of the curve associated with a physical system leads to the concept of its equation of \emph{state}, contrasting with the traditional equation of \emph{motion}. Most importantly, the Mach-Poincar\'e principle can be modernized to reflect the intrinsic geometric nature of PSD:

\setcounter{tenn}{0}

\begin{tenn}[\textbf{Mach-Poincar\'e Principle -- Modern Version}]
\label{MachPSD}
Physical, i.e., relational initial configurations and their (intrinsic) derivatives alone should uniquely determine the dynamical evolution of a closed system.
\end{tenn}

The innovation of revised tenet \ref{MachPSD} lies in considering \emph{higher-order} derivatives of the curve, allowing the description of a system’s dynamics purely through the curve itself, without additional non-shape parameters. While revised tenet \ref{MachPSD} requires more initial data than tenet \ref{MachP}, it achieves a fully relational dynamics by eliminating any degree of freedom external to the system.

For a quotiented out physical system, the equation of state of the unparametrized curve $\gamma_0$ in shape space $\mathcal{Q}_{ss}$ can be schematically expressed as:

\begin{equation}
\begin{array}{rcl}
dq^a&=&u^a(q^a,\alpha _I^a), \\
d\alpha _I^a &=&A _I^a(q^a,\alpha_I^a).
\end{array}
\label{curve0}
\end{equation}

The right-hand sides must be described in terms of dimensionless and scale-invariant quantities, whose intrinsic change is obtained using Hamilton's equations of motion. Here, $q^a$ are points in shape space representing universal configurations of the system, and $u^a$ is the unit tangent vector defined by the shape momenta $p_a$:
\begin{equation}
u^a\equiv g^{ab}(q)\frac{p_b}{\sqrt{g^{cd}p_cp_d}}.
\label{unittangent}
\end{equation}
This vector allows the definition of the direction $\phi^A$ at $q^a$ ($g^{ab}$ being a naturally defined metric over shape space; see \S~\ref{sec:plat} below). The shape momenta enter Hamilton's equations via the unit tangent vector and the associated direction, which are used in intermediary steps leading to the equation of state \eqref{curve0}. Finally, $\alpha _I^a$ represents any additional degrees of freedom needed to fully describe the system, including higher-order derivatives of the curve (which get rid of the need of additional non-shape degrees of freedom, unlike standard
SD), and $A _I^a$ is a compact way to write the set of functional relations accounting for the evolution of said degrees of freedom. It is important to stress that the equation of state \eqref{curve0} should be taken as a whole and interpreted as giving the relative rates of change of the degrees of freedom of the curve in shape space. This means that $\frac{dq^a/ds}{d\alpha _I^a/ds}=\frac{dq^a}{d\alpha _I^a}$ (the arc-length parameter $s$ dropping out), which renders the dynamics explicitly unparametrized.

The unifying nature of \eqref{curve0} is notable: It encapsulates the entire framework of relational dynamics---classical, relativistic, and quantum. In a nutshell, the more ``complicated'' is the system to be described, the more intrinsic degrees of freedom $\alpha _I^a$ will appear in the equation of state. Here lies the fundamental difference between standard SD and PSD. Indeed, PSD relies solely on the intrinsic geometric properties of the dynamical curve in shape space, whereas standard SD does not emphasize this aspect. The unparametrized character of the curve in PSD, ensuring no external references or clock processes are required, highlights this intrinsic property focus. 

Despite this intrinsic nature, PSD can reproduce known physics, as demonstrated in \citet[][appendix A]{746} for the $E=0$ $N$-body problem. Most notably, PSD constrains by construction the total angular momentum of the universe $\mathbf{J}$ to be zero, pointing to the fact that any scenario with $\mathbf{J}\neq0$ should be regarded as unphysical---thus defusing the issue brought up by Poincar\'e. This also eliminates at the root any need to appeal to absolute space to define a privileged frame of reference to make sense of inertial effects, thus dissolving the problem posed by Newton's bucket argument. In PSD, inertial effects like the concavity of the water in a spinning bucket are explained by considering the relations among all the bodies in the universe. These relations are subject to the condition that the total angular momentum of the universe is zero. This constraint realizes Mach’s famous idea that the concavity arises because the water rotates relative to the entire material content of the universe---not just relative to the bucket’s walls. Thus, rather than needing an absolute space to account for inertial effects, PSD shows how they emerge naturally from the universe’s global relational structure.

Another critical aspect of PSD is its approach to dynamics. Without parametrized curves, PSD seeks an intrinsic feature of the system for a meaningful labeling of change. Following the idea proposed in \cite{706}, the evolution of a system is directed towards configurations that maximize its \emph{complexity} (e.g., the amount of clustering among particles in the Newtonian case). This provides a basis for introducing a direction of change in terms of the accumulation of stable records---that is, subsystems that maintain their ordered structure throughout the dynamical development.

PSD also recovers standard dynamics for subsystems of the universe. In the classical $N$-body case, the dynamics leads to the formation of individual particles and clusters that become increasingly isolated in the asymptotic regime \citep{saar}. These subsystems develop approximately conserved charges like energy $E$, linear momentum $\bf P$, and angular momentum $\bf J$. Within these subsystems, pairs of particles can function as physical rods and clocks, often referred to as \emph{Kepler pairs} due to their asymptotic elliptical Keplerian motion. In other words, the conserved charges developed by these subsystems make it possible to define units of scale $X^2=\frac{\bf J^2}{\bf P^2}$ and duration $T^2=\frac{\bf J^2}{E^2}$, thus recovering the usual spatial and temporal notions employed in standard physics \citep[see][for a thorough discussion of how standard temporal notions emerge in the classical $3$-body model of PSD]{763}.

In conclusion, PSD represents a natural evolution of Barbour and Bertotti's ideas, providing a robust formal framework for a comprehensive ``Machianization'' of physics. While much work remains to be done, the technical foundations of the theory are already well-established.

\section{Gaining physical knowledge in a spacetimeless world}\label{sec:two}

In the PSD framework, there is clearly no possibility of maintaining an ontic commitment towards space and time as self-subsisting entities. This is because, for example, in the relational $N$-body case, the fundamental picture of the dynamics is \emph{not} that of a set of $N$ particles changing position in an external Euclidean space to the ticking of a universal clock. In PSD, the fundamental dynamical structure of a $N$-body model is that of a timeless sequence---encoded in an unparametrized curve---of ``instantaneous'' universal relational configurations (i.e., shapes), each configuration being just a set of $N$ \emph{relata} interconnected by a web of conformal Euclidean relations. Such a sequence is timeless in that all shapes are given ``at once,'' so to speak---there is no fundamental sense in which a configuration comes ``before'' or ``after'' another. The dynamics is also spaceless because it does not need to be embedded in an external space to be traced: Each relational configuration is characterized and individuated in purely intrinsic terms, without the need to be placed anywhere. Otherwise said, according to PSD, what makes the dynamics intelligible to our minds is not space and time, but the intrinsic properties of a system. This, of course, does not mean that the framework is unable to recover a spacetimeful description of physics at a certain point, as we have seen in the previous section. However, the full spatial and temporal description of physics recovered from PSD is just a simple and informative way to describe a ``local'' perspective on the dynamics of the subsystems within the universe---the fundamental, objective structure of the world remains holistic and spacetimeless. Obviously, as already mentioned in the previous section, this discussion applies beyond the $N$-body model of PSD---any closed system whose symmetries can be quotiented out are amenable to the Leibnizian/Machian interpretation at the root of the PSD framework. 

To sum up, PSD's perspective on dynamics vindicates the relationalist denial of the substantivalist ontic commitments toward space and time, while supplying a story about how spatial and temporal notions can be recovered in ``everyday'' parlance. However, remembering what has been said in \S~\ref{sec:int}, this means that PSD meets \emph{half} of the challenge raised against anti-substantivalists. The problem is now shifted to clarifying how the PSD framework accounts for the ``relational intelligibility'' of the world. Answering this question means finding the appropriate interpretation of shape space---i.e., the ``arena'' where the dynamics takes place---which enables a viable physical representation. Note that just claiming that shape space is a relational state space, which serves the same purpose as standard configuration space in Newtonian physics, would cut no ice. Indeed, standard configuration space acquires the ``power'' to represent physical states of affairs through a pre-existing commitment to Newtonian space and time: This is why a point in configuration space is not just an $n$-tuple of real numbers, but a specification of how each element of the physical system is placed in space and a given time. Obviously, such a story cannot be provided in the shape space case. Let's consider two possible solutions to this problem in more detail and, to keep things simple, let's focus on the shape space of a Newtonian equal-mass 3-body model of PSD.


\subsection{Shape space as a physical space}\label{sec:plat}

As previously mentioned, shape space is the relational configuration space of a system, derived from ordinary configuration space by quotienting out (in the case of Newtonian particles) the similarity group of translations, rotations, and dilations. In compliance with the relational ideas posited by Leibniz and Mach, we focus on systems with zero total energy $E$, zero total linear momentum $\mathbf{P}$, and zero total angular momentum $\mathbf{J}$---constraints that naturally emerge from the tenets discussed in \S~\ref{sec:one}. Particularly, the zero $\mathbf{J}$ condition has an immediate consequence for the 3-body system: It becomes planar. This planar nature of the 3-body system leads to a significant topological insight, namely, the configuration space of the 3-body system has the topology of a sphere \citep{755,mont}. Consequently, its associated shape space is aptly termed the \emph{shape sphere} (\citealp{712}, \S~2.6; \citealp{756} \S~3). Without going into technical details, the $3$-body problem is first rewritten in translationally invariant coordinates, dubbed \emph{Jacobi coordinates}; then, these coordinates are appropriately combined to form a new set of coordinates $\{w_{1},w_{2},w_{3}\}$ that additionally show rotational invariance. These are called \emph{Hopf coordinates} and constitute a complete coordinate system over the scale-dependent relational space. By setting $\lVert\mathbf{w}\rVert^{2}=1$ (with $\mathbf{w}=(w_{1},w_{2},w_{3})$) it is possible to eliminate any scale dependence and thus ``project'' this relational space down to shape space---which, hence, is just the unit sphere in the $(w_{1},w_{2},w_{3})$-space.

\begin{figure}[h!]
\begin{center}
\includegraphics[scale=1.5]{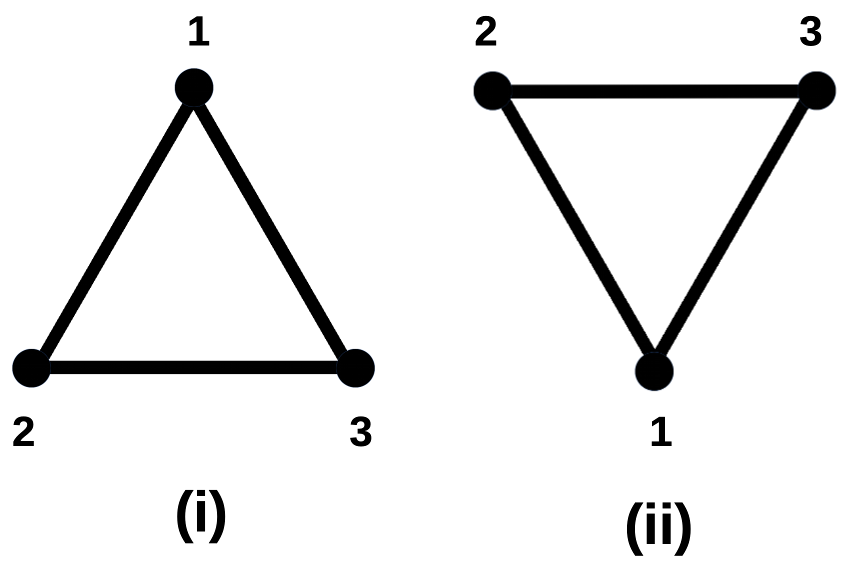}
\caption{Operation of mirror conjugation in the simple case of an equilateral configuration. Note how it is not possible to get from (i) to (ii) by a rotation over the Euclidean plane. This is why (i) and (ii) can be considered distinct configurations on shape space provided the three particles are distinguishable.}
\label{fig2}
\end{center}
\end{figure}
\begin{figure}[h!]
\begin{center}
\includegraphics[scale=2]{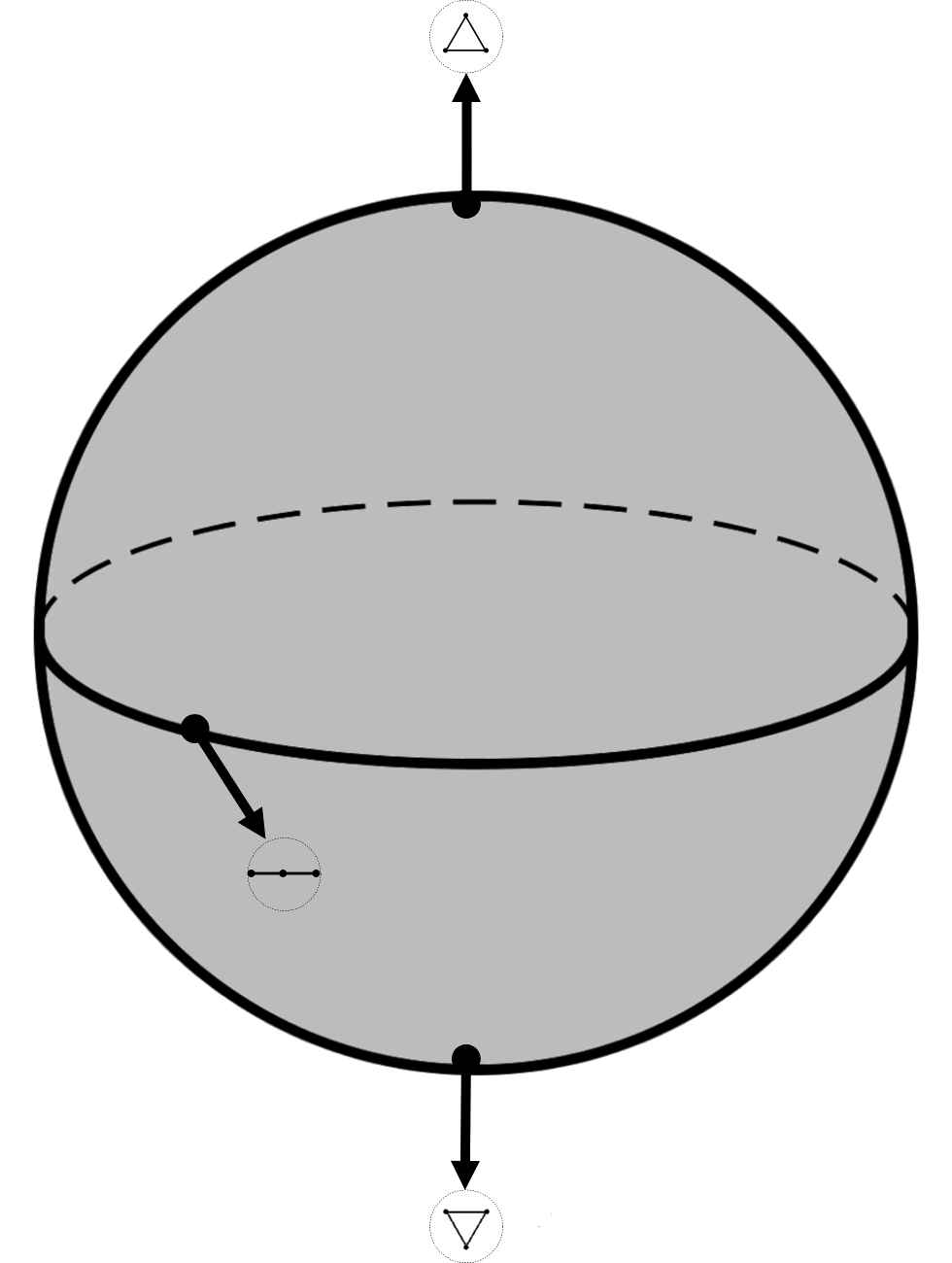}
\caption{Schematic representation of the shape sphere for the equal-mass $3$-body problem. Each point on the sphere's surface represents three particles arranged in a triangular shape. In particular, the points at the pole represent the  equilateral triangle shape, while the points on the equator represent collinear configurations (degenerate triangles with null area). Tracing a continuous curve connecting the three points shown in the figure would represent a dynamical evolution of an equilateral triangle getting progressively ``squashed'' into a line and then returning to its starting (non-oriented) shape \citep{klin}.}
\label{fig1}
\end{center}
\end{figure}

On this shape sphere, every point represents a specific triangle. Indeed, the shape sphere is the set of oriented similarity classes of triangles. This means that points sharing the same longitude but having opposite latitudes correspond to mirror-conjugated triangles. Note that the operation of mirror conjugation does \emph{not} correspond to a global planar rotation (which would count as a redundant transformation, see figure \ref{fig2}). This distinction between ``clockwise'' and ``counterclockwise'' configurations facilitates the shape space treatment of the Newtonian $3$-body problem in that a curve can smoothly pass through the equator (otherwise this circle would be a singularity of this space) but, in principle, mirror-image configurations can be identified if we take the particles to be totally indistinguishable. This, however, would alter the topology of the space \citep[see, e.g., the detailed technical discussion in][\S~2.3]{446}. At the poles of this sphere lie the equilateral triangles, while the equator hosts the collinear configurations, which are degenerate triangles with null area  (see figure \ref{fig1}). A crucial aspect to note is the nature of the metric induced on shape space, which already appeared in \eqref{unittangent}. This metric measures the degree of \emph{similarity} between shapes rather than distances between them: The ``closer'' two shapes, the more similar. Thus, a continuous curve traversing the shape sphere does not merely represent a sequence of locations but an ordered sequence of triangular shapes, each differing increasingly in their intrinsic geometrical properties---i.e., the internal angles.

How should this characterization of the shape sphere be interpreted in physical terms in order to address the problem of accounting for the intelligibility of the dynamics without invoking a spatiotemporal ``arena'' in which the dynamics unfolds? A straightforward answer to this question is to claim that such an ``arena'' for the dynamics is the shape sphere itself. On this view, a curve solution to \eqref{curve0} represents the dynamics of a system in that the physical information encoded in \eqref{curve0} \emph{literally} constitutes a path over the shape sphere, i.e., a sequence of shapes over the shape sphere. In this sense, a dynamical curve in PSD would represent the spacetimeless equivalent to a spatiotemporal trajectory in non-relational physics. According to this perspective, which we might call \emph{shape space realism}, every point---and every sequence of points---in shape space must be regarded as ontologically equal. In other words, shape space realism claims that all triangular shapes are real and actual, making shape space the complete collection of all universal relational configurations \emph{simpliciter}---exactly like Newtonian space is the collection of all spatial locations. For shape space realists, shape space is the actual world itself. Note that this notion is reminiscent of some ideas predating the development of the shape dynamics program. Indeed, Julian Barbour famously termed this relational world \emph{Platonia} \citep{10}. 


Shape space realism vindicates relationalism's empiricist spirit, according to which a viable representational framework for physics should be grounded in experiential data---i.e., the conformal structure inherent into a shape---rather than pre-existing spatial and temporal notions. It may be objected that, at first glance, the conformal structure inherent in a shape—such as the internal angles characterizing a triangular particle configuration—may seem less immediately tied to experience than directly measurable, operational properties like length or duration. However, from the relationalist perspective defended here, the apparent experiential immediacy of operational properties actually rests on deeper relational facts. Physical measurements, while often framed using rulers and clocks under specific measurement unit systems, derive their significance from the invariances they reveal: The stable ratios, angles, and relative arrangements that remain unaffected by shifts, rotations, or scalings of the reference frame. In short, once we strip away the scaffolding of substantival space and time, what genuinely carries empirical and experiential weight are the invariant relational patterns---the conformal structures---that persist regardless of measurement conventions. These relational features are not mere abstractions but are precisely what observers track, however implicitly, when they compare, measure, and interpret the physical world. Thus, conformal structure offers not only a parsimonious metaphysical description but also the clearest account of what is empirically accessible when we interface with the external world.

By providing a direct metaphysical link between the formalism of PSD and a timeless physical world of shapes, shape space realism neatly explains how such a world is intelligible to us: Simply, through measurements, intended as comparisons between physical magnitudes (in the relational $3$-body case, just length comparisons). Hence, it seems that shape space realism succeeds in dispensing with substantivalism's ontic commitments without undermining our reasons to believe that the physical world can be known. There is, however, quite a price to be paid to reach this result.

As we mentioned in section \ref{sec:int}, one of the biggest complaints relationalists had against substantivalists is that they postulate \emph{too much} structure to make sense of physics. But, now, it seems that shape space realists are committing the same sin, if not worse. Indeed, while shape space realism dispenses with space and time, it basically substitutes it with an even larger structure, i.e., shape space itself. One may point out that shape space realists do not face the same redundancy problem encountered by substantivalists, as the whole of shape space consists of empirically distinguishable configurations---meaning there cannot be cases of ontologically distinct yet empirically equivalent states of affairs. However, this is not entirely true. To see this, consider our universe: Its entire history of observable relational changes among the material entities inhabiting it is encoded in a \emph{single} curve on shape space\footnote{Of course, a much more complicated one than the shape sphere in figure \ref{fig1}.} generated through \eqref{curve0} given certain initial conditions. All the other curves spanning shape space seem to be \emph{possible} physical evolutions that might have ensued given different initial conditions. Note how this reading introduces a modal ``asymmetry'' into the picture: How the universe actually is, is obviously distinct from the other ways the universe might have been. In empiricist terms, this point can be put like this: Physical possibility is not observable, only physical actuality is. This is troublesome for shape space realism because, by definition, it denies that there is any modal difference between shapes---they are all actual---and, hence, it cannot account for the modal asymmetry that we experience. While there may be a way out of this issue by claiming some mechanism through which we get to experience some relational configurations but not others---Barbour called these experientially privileged configurations ``time capsules'' \citep{136}---it is legitimate to ask whether an alternative interpretation of shape space can be provided, which sidesteps the issue. In the remainder, it will be argued that the answer is a resounding \emph{yes}: Shape space, far from being a \emph{sui generis} physical entity, is just a particular case of a conceptual space.

\subsection{Shape space as a conceptual space}\label{sec:scon}

In the realm of cognitive science and philosophy, the framework of conceptual spaces, as proposed by Peter G\"ardenfors \citep{761}, offers a compelling bridge between symbolic and connectionist approaches to understanding human cognition. The symbolic approach posits that cognitive systems function like Turing machines, where cognition is essentially computation involving the manipulation of symbols. On the other hand, connectionism maintains that cognitive representations arise from associations among different information elements, which can be modeled with artificial neural networks. However, there are certain aspects of cognitive phenomena that neither approach seems to adequately address. One significant challenge is the mechanism of concept acquisition, which is crucial for understanding many cognitive processes. Concept learning is closely tied to the notion of similarity, something that proves problematic for both symbolic and connectionist models. As a result, neither approach provides a fully satisfactory framework for explaining how concepts are learned and represented in the mind. Against this background, the conceptual spaces framework provides a middle ground that captures the nuances of conceptual representation often lost in purely symbolic or neural network models.


At its core, the conceptual spaces framework posits that concepts are best understood as regions within multi-dimensional spaces. These conceptual spaces are constructed by ``combining'' quality dimensions, which can be thought of as axes along which properties of objects are measured. Quality dimensions are foundational to understanding conceptual spaces. They can be simple, like temperature or weight, or more complex, like taste or emotional valence. These dimensions are often perceptual but can also include more abstract qualities like moral goodness or social status. To illustrate the practical application of the conceptual spaces framework, let's consider the concrete case of color perception and categorization. Color provides an excellent example because it is a perceptual quality that can be measured along several dimensions. Indeed, within the framework of conceptual spaces, color can be represented in a three-dimensional space with the following quality dimensions:

\begin{enumerate}
\item \emph{Hue}, corresponding to the type of color.
\item \emph{Saturation}, indicating the intensity or purity of the color.
\item \emph{Brightness}, reflecting the lightness or darkness of the color.
\end{enumerate}

These three dimensions form a conceptual space where every possible color can be located at a specific point. For example, the color sky blue would have particular values along the hue, saturation, and brightness dimensions. Within this color space, different colors shades are represented as regions. A region is defined by a cluster of points that share common properties. Consider the concept of ``blue.'' This concept does not refer to a single point but to a region that includes various shades of blue, from navy to sky blue. The prototype concept of blue might be a specific shade that people generally agree represents the most typical example of blue, such as a medium blue that is neither too dark nor too light. This prototype serves as a reference point within the conceptual space, helping individuals categorize and identify new instances of blue. When a new color is encountered, it is categorized by its proximity to existing regions in the color space. If someone sees a new shade of blue-green, they locate this color within the hue, saturation, and brightness dimensions and determine its nearest region. If it is closer to the ``blue'' region than to the ``green'' region, it is likely to be categorized as a type of blue. This process illustrates how conceptual spaces allow for flexible and context-sensitive categorization. Instead of rigidly defining colors with strict boundaries, the regions within the conceptual space overlap, reflecting the continuous nature of color perception. 

Conceptual spaces also account for variations across different languages and developmental stages. In some languages, there may be no distinct word for ``blue'' and ``green,'' instead using a single term like ``grue.'' In this case, the conceptual space for colors in these languages would have a broader region that encompasses both what English speakers would categorize as ``blue'' and ``green.'' Moreover, as individuals gain more experience with colors, their conceptual spaces can evolve. Consider a child learning about colors. Initially, they might have broad regions with few distinctions (e.g., all shades of red might be categorized simply as red). As they learn more and refine their perceptions, these regions become more nuanced, differentiating between crimson, scarlet, and burgundy, and thus allowing for more precise and detailed categorization. This dynamic aspect of conceptual spaces reflects how learning and experience shape our understanding of concepts. Given this analytical power and flexibility in modeling human cognition, it is then no surprise that the framework of conceptual spaces has profound implications for various fields beyond cognitive science, including artificial intelligence, education, and marketing \citep{garzen}.



The conceptual spaces framework suggests an immediate solution for making epistemic sense of shape space in PSD without reifying it. To see this, it is sufficient to think of applying the same construction carried out for the color case to the problem of classifying triangular shapes in Euclidean space. Just as with color perception, the cognition and classification of triangles can be effectively modeled within a multi-dimensional conceptual space. How does this space look like? \emph{Prima facie}, it should contain three quality dimensions, each corresponding to an internal angle of a triangle. This characterization is, however, too naive in that these three dimensions are obviously not independent---the sum of the internal angles have to be $180^\circ$. Hence, in order to construct such a conceptual space in a mathematically rigorous way, we should first take the configuration space of all possible triples of points on the Euclidean plane and then identify all configurations related by a rigid translation, rotation, or scaling. But this is exactly the procedure we have already encountered to construct the shape sphere in figure \ref{fig1}! 

Indeed, within the shape sphere, different regions correspond to different categories of triangles. For example, the concept of ``isosceles triangle,'' which consists of triangles with two equal angles, corresponds to three great circles, each intersecting both poles (i.e., the equilateral configurations) and the equator at one of the three so-called \emph{Euler configurations} (which, very roughly speaking, correspond to the collinear configurations where the particles are equally spaced).\footnote{If the reader is having a hard time figuring out why there are three great circles for isosceles triangles, they can pictorially think that there are three ways in which an equilateral configuration of three particles can be ``squashed'' into an equidistant collinear configuration by letting one of the particles ``move down'' toward the line of the other two held fixed (remember that, on the shape sphere construction, the particles are assumed to be distinguishable, so each choice is a separate case).} A prototype for an isosceles triangle might be one where the angles form a ``very regular'' shape such as $45^{\circ}-45^{\circ}-90^{\circ}$. This prototype aids in the classification of new triangles by serving as a benchmark within the conceptual space---e.g., when teaching a child what an isosceles triangle is. Moreover, when a new triangular shape is encountered, it is categorized by comparing its angles to the regions within the triangle space. For example a triangle with angles $43^{\circ}-62^{\circ}-75^{\circ}$ would be classified as being ``less regular'' than a $50^{\circ}-45^{\circ}-85^{\circ}$ triangle based on its greater distance from the region representing isosceles triangles. Note how the example mirrors the flexibility seen in color categorization. Just as color regions overlap, allowing for nuanced classifications, some regions on the shape sphere may overlap, accommodating the continuum of shapes between distinct types of triangles. The bottom line of this analysis is evident: \emph{The shape sphere is a clear case of a conceptual space.}

That shape spaces are closely related to conceptual spaces is no bold thesis. Indeed, the two frameworks share key characteristics. First, both approaches use multidimensional spaces for representation: In conceptual spaces, dimensions represent perceptual or conceptual attributes, while in shape spaces, dimensions correspond to different aspects of spatial relations. Second, both frameworks have a characteristic geometric structure that orders objects in terms of their similarity. Third, and most importantly, both frameworks emphasize grounding representations in perceptual experience, including empirical observations and physical measurements. Indeed, the original proposals for shape space representations came from the need to provide a rigorous account of objects' shape and shape change amenable to mathematical and statistical analysis in the context of many different fields of study, including biology, archaeology, and astronomy \citep[see, e.g.,][]{ken,boo}. In short, shape space does not just model the intrinsic dynamics of physical systems. It also models the way we get to know and understand such a dynamics through empirical investigation. And all of this with fewer ontic commitments than substantivalism requires. 

Note also that the conceptual clarity and flexibility that comes from equating shape space to a conceptual space avoids all the drawbacks faced by shape space realism. For example, abandoning the presupposition that particles are distinguishable would not have major consequences if we embrace the view that shape space is a conceptual space: We should just have to change the topology of shape space to reflect this change in representation. Instead, for a shape space realist, accepting the indistinguishability of particles would radically alter their ontology in that it would mean abandoning the commitment to the physical world being a shape sphere in favor of a radically different structure. Moreover, the concern regarding modal asymmetry---that is, the apparent difference between the actual trajectory of the universe through shape space and the mere potential trajectories---is naturally avoided under the conceptual space reading. The idea is that the trajectory corresponding to the actual universe is distinguished by the fact that it embodies the evolving relational configuration we observe---trivially, we directly experience what is actual, not what is possible. Thus, the conceptual space framework allows us to understand the modal asymmetry as a difference in how shapes are situated within our epistemic perspective. Hence, by emphasizing shape space's role as a representational tool rather than a metaphysical arena populated by distinct modal realities, we can coherently account for the asymmetry without compromising the relationalist thesis.

In the end, by accepting that shape space is a conceptual space, it is easy and simple to justify the claim, central to Leibnizian/Machian relationalism, that we get to know the physical world through its intrinsic representation, i.e., the web of ever-changing conformal relations among material bodies.

\section{Conclusion: Beyond classical physics}\label{sec:disc}

The discussion in this paper has been carried out in the context of Newtonian gravity. This immediately prompts the question as to whether the ``shape space as a conceptual space'' narrative holds water beyond this restricted context. This question can be certainly answered in the positive as far as classical physics in general is concerned. Indeed, we could repeat the same reasoning in \S~\ref{sec:scon}, \emph{mutatis mutandis}, for the case of continuum mechanics and classical field theory. In these cases, the shape space geometries would obviously be much more complicated than that of the shape sphere. For example, the shape space of classical field theory would feature all the possible relational configurations of field intensities \citep[see, e.g.,][\S~2.1]{724}. However, despite this increase in mathematical complexity, these spaces could still be regarded as conceptual spaces encoding the perceptual qualities associated with interacting with such fields or material \emph{continua} upon measurement.

What about the case of general relativity (i.e., dynamical geometry) and quantum physics? At the moment, the extension of PSD to these domains is work in progress (see, e.g., \citealp{732}, \S~2.4.4, and \citealp{fark}, for some preliminary results) but some considerations can already be made. As regards the case of dynamical geometry, the shape space would be the result of quotienting volume preserving conformal transformations out of the space of $3$-geometries---which is called \emph{superspace}, and represents the configuration space of general relativity \citep[][chapter 43]{27}. Despite the extreme complexity of this ``reduced''  superspace, still the case can be made of it representing a conceptual space: Roughly speaking, it can be taken as a ``conceptual map'' that can be used to categorize and compare conformal $3$-geometries. The case of quantum physics is much more challenging, given that it requires a clear-cut answer to the question as to whether---and if yes, how---the wave function should be accounted for by a set of quality dimensions. The best case scenario would of course be one where it could be shown that the wave function just encodes physical properties that can be ascribed to material systems, since this would mean that the shape space of the theory would be identical to the classical case---i.e., one in which the perceptual qualities stem from the relative placement of material bodies. However, such an ideal scenario seems impossible to implement, even in the very simple case of the relational $N$-body de Broglie-Bohm theory \citep[][\S~4]{757}, meaning that the challenge of finding a place for the wave function into the conceptual space interpretation of PSD is inescapable. Note how this challenge is deeply related to the problem of coming up with an appropriate ontology of the quantum world. Hopefully, progress in developing a Leibnizian/Machian framework for quantum physics will bring conceptual clarity to the table.

\pdfbookmark[1]{Acknowledgements}{acknowledgements}
\begin{center}
\textbf{Acknowledgements}:
\end{center}
I am grateful to the Editors of this topical collection for their kind invitation to contribute. Many thanks also to Pedro Naranjo and two anonymous referees for their comments on a previous version of the manuscript. Finally, I gratefully acknowledge financial support from the Polish National Science Centre, grant No. 2023/49/B/HS1/01091.


\bibliography{biblio}

\end{document}